\documentclass[doublespacing]{elsart4}
\usepackage[latin1]{inputenc} 
\usepackage[english,francais]{babel} 
\usepackage{graphicx,epsfig,color} 
\usepackage{amsmath,amssymb} 
\usepackage{array,hhline} 
\usepackage{fancyheadings} 
\renewcommand{\theequation}{\arabic{equation}}
\begin{document}

\centerline{Astrophysique/Système Solaire}

\begin{frontmatter}

\selectlanguage{francais}

\title{Formes d'astéroïdes et formation de satellites : rôle de la réaccumulation gravitationnelle}

\author[tanga,consigli]{Jean-François Consigli}
\author[tanga]{Paolo Tanga} 
\author[tanga,comito]{Carlo Comito}
\author[hestroffer]{Daniel Hestroffer} 
\author[richardson]{Derek C. Richardson}
\address[consigli]{Université de Nice-Sophia Antipolis, France}
\address[tanga]{Observatoire de la Côte d'Azur, Nice, France}
\address[hestroffer]{IMCCE, Observatoire de Paris, France}
\address[richardson]{Université du Maryland, USA}
\address[comito]{Universit\'a di Pisa, Italy}


\begin{abstract}
\selectlanguage{francais}
Plusieurs indices suggèrent qu'un grand nombre d'astéroïdes seraient des empilements de gravats, liés uniquement par gravité et quasiment dépourvus de cohésion interne. Leur formation serait due à la réaccumulation de fragments issus de la destruction antérieure d'un corps parent. La forme de ces objets, que l'on observe, pourrait ainsi être liée à ce processus de réaccumulation gravitationnelle. Toutefois, à l'heure actuelle, on ignore si les formes observées des astéroïdes sont le résultat de la seule réaccumulation ou d'événements ultérieurs entrainant un réajustement de la forme du corps.\\
Dans cet article, nous nous interrogeons sur l'origine des formes des astéroïdes. A l'aide d'une étude numérique de la réaccumulation, nous étudions les formes d'équilibres possibles correspondants aux modèles théoriques. Les résultats montrent, pour la première fois, que seulement une catégorie précise de formes (sphéroïdes aplatis) est apparemment crée via ce mécanisme. Ces résultats pourraient ainsi fournir d'interessantes contraintes sur l'évolution des formes d'astéroïdes, notamment pour ceux possédant un, ou plusieurs, satellites.\\
{\it Pour citer cet article~: J.-F. Consigli et al., C. R. Physique ... (2006).}
\vskip 0.5\baselineskip
\selectlanguage{english}
\noindent{\bf Abstract}
\vskip 0.5\baselineskip
\noindent
{\bf Asteroid shapes and satellites: role of gravitational reaccumulation.}
Following current evidences, it is widely accepted that many asteroids would be "gravitational aggregates", i.e. bodies lacking internal cohesion. They could mainly be originated during the catastrophic disruption of some parent bodies, through the gravitational reaccumulation of the 
resulting fragments. The same events produced the dynamical families that we observe. In this work we address the problem of the origin of shapes of gravitational aggregates, that could contain signatures of their origin. We use a N-body code to simulate the collapse of a cloud of fragments, with a variety of initial velocity distributions and total angular momentum. The fragments are treated as inhelastic spheres, that rapidly accumulate to form rotating aggregates. The resulting shapes and rotational properties are compared with theoretical predictions.
The results show that only a precise category of shapes (flattened spheroids) are created via this mechanism. This may provide interesting constraints on the evolution of asteroid shapes, in particular for those with one or more satellites.\\
{\it To cite this article: J.-F. Consigli et al., C. R. Physique ... (2006).}

 \vskip 0.5\baselineskip
\noindent{\small{\it Mots-cl\'es~:} Système Solaire; Astéroïdes; Satellites; Collisions}}
\vskip 0.5\baselineskip
\end{abstract}
\end{frontmatter}

\selectlanguage{francais}



\section{Introduction}

Depuis leur formation, tous les astéroïdes évoluent principalement
sous l'influence des multiples perturbations planétaires et des chocs
qu'ils subissent entre eux.
Plusieurs études \cite{Davis},\cite{Farin2} montrent 
qu'hormis les plus grands d'entre eux (Cérès, Pallas, Vesta, Juno...), les astéroïdes seraient le
résultat de collisions mutuelles.\\
Parmi ces collisions, les plus violentes d'entre-elles, appelées collisions
catastrophiques, 
sont responsables de la destruction totale du corps impacté et génèrent ainsi
des centaines de milliers de fragments dont il demeure, encore aujourd'hui, des
traces observationnelles. Le témoignage de cette intense évolution
collisionnelle est apporté par des groupes d'objets partageant des
compositions et des propriétés orbitales similaires. L'existence de ces
groupes d'objets, les familles dynamiques d'astéroïdes, est désormais
clairement établie \cite{Farin2}. A ce jour, une vingtaine de ces familles a 
ainsi été découverte dans le Système Solaire.\\
\\
Contrairement à ce que l'on pensait jusqu'alors, on sait depuis peu \cite{Michel} que
la formation d'une famille d'astéroïdes ne peut s'expliquer uniquement par la
fragmentation 
d'un corps ``parent". Dès qu'un objet dépasse
quelques centaines de mètres, sa destruction signifie qu'il se
fragmente lors de la traversée de l'onde de choc, mais également que
les fragments produits s'échappent et interagissent mutuellement. Certains fragments peuvent se réaccumuler
sous leur propre attraction pour former des agrégats qui
constitueraient ainsi les membres des familles d'astéroïdes
observées aujourd'hui \cite{Michel}.\\
Outre la formation de ces familles, ces collisions peuvent également entraîner
la formation de satellites \cite{Durda96},\cite{Durda03} 
en orbites stables \cite{Chauv},\cite{Schee2} autour d'astéroïdes. Pratiquement 
dépourvus de toutes forces de cohésion interne, ces corps réaccumulés
seraient des agrégats de fragments liés uniquement par la
gravité mutuelle s'exercant entre les différents fragments constituant l'agrégat lui-même.\\
\\
Parallèlement à ces travaux, des preuves indirectes de l'existence de
ces agrégats ont été obtenues. On peut citer à ce propos les études de courbes
de lumière \cite{Pravec}, \cite{Harris}, ou les observations
de l'astéroïde Mathilde par la sonde NEAR.
Mathilde possède en effet une faible densité, $\rho=$1.3 g cm$^{-3}$, soit
environ 3 fois inférieure aux astéroïdes de composition similaire
\cite{Yeo}. Un tel constat laisse penser que Mathilde serait en réalité un
corps réaccumulé, de grande porosité (pourcentage de vide), formé de débris
issus d'une collision antérieure.\\
L'objet de ce travail est précisément d'examiner les caractéristiques des
astéroïdes se formant par réaccumulation gravitationnelle lors de
l'effondrement d'un nuage de particules. La question centrale que l'on se pose
est de savoir quels liens il existe entre la forme des astéroïdes, la présence
éventuelle de satellites et le processus de réaccumulation lui-même. Afin
d'apporter des éléments de réponses, on réalise une étude numérique ainsi
qu'une comparaison avec des modèles de formes d'équilibres déjà existants.
L'approche utilisée au cours de ce travail se caractérise par son aspect
novateur. Une étude précédente \cite{Rich} prenait en effet comme point de
départ des agrégats \textit{déjà formés} et s'intéressait à la façon dont la
forme de ces corps évoluait et se réajustait en fonction de leur vitesse de
rotation. La démarche utilisée ici est différente : on souhaite comprendre
quelles sont les formes ``possibles" (accessibles) via le processus de
réaccumulation, et ce, sans faire appel à un corps déjà formés, mais en
considérant uniquement l'effondrement d'une distribution dispersée de
particules, censée représenter un nuage de débris après une collision catastrophique.\\
\\
Nous évoquerons dans la partie~2 les différents modèles de formes
d'équilibres. Dans la partie~3, nous discuterons de la méthode
numérique utilisée pour mener à bien les différentes simulations. Enfin, en~4, nous analyserons les résultats issus de ces mêmes simulations
afin d'en tirer des conclusions préliminaires quant à l'origine des formes des astéroïdes que l'on observe.

\section{Formes d'équilibre}

\begin{figure}
   \resizebox{\hsize}{!}{
     \includegraphics[angle=0]{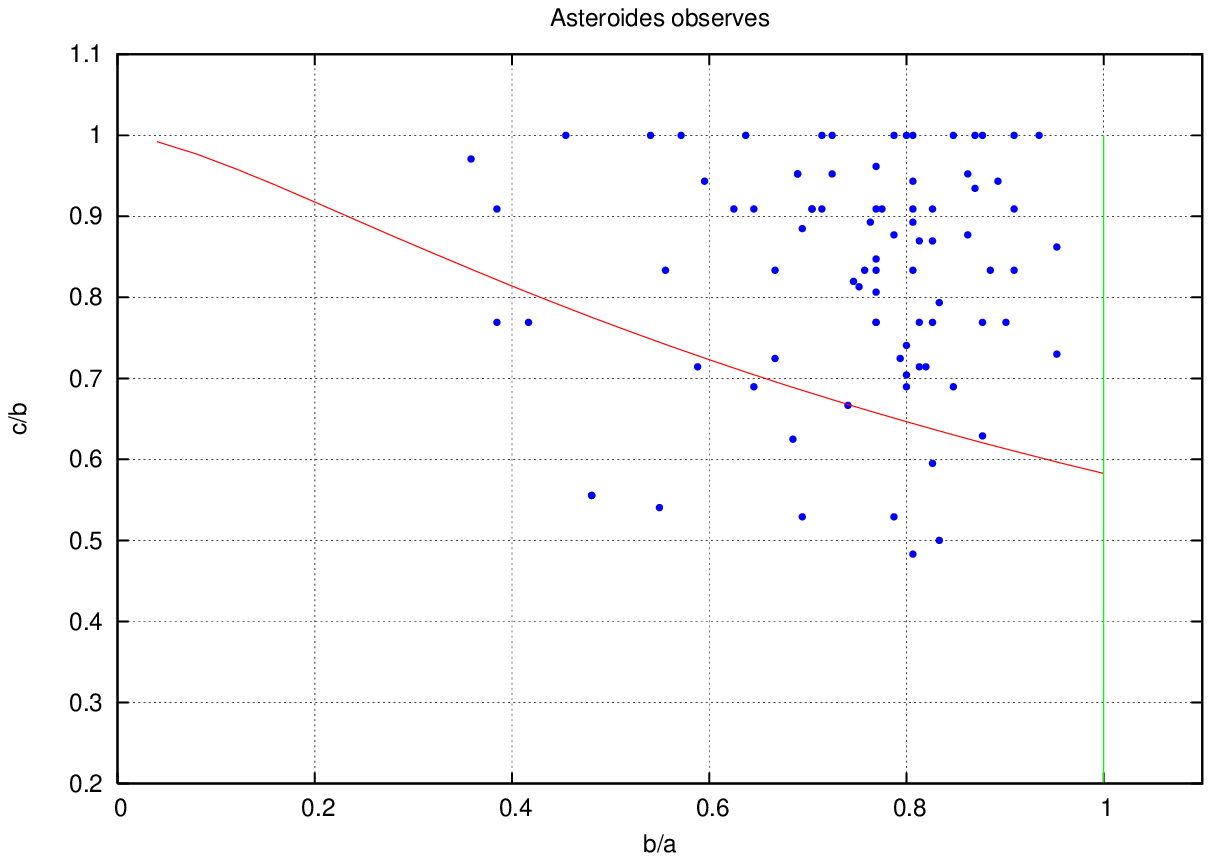}
     \includegraphics[angle=0]{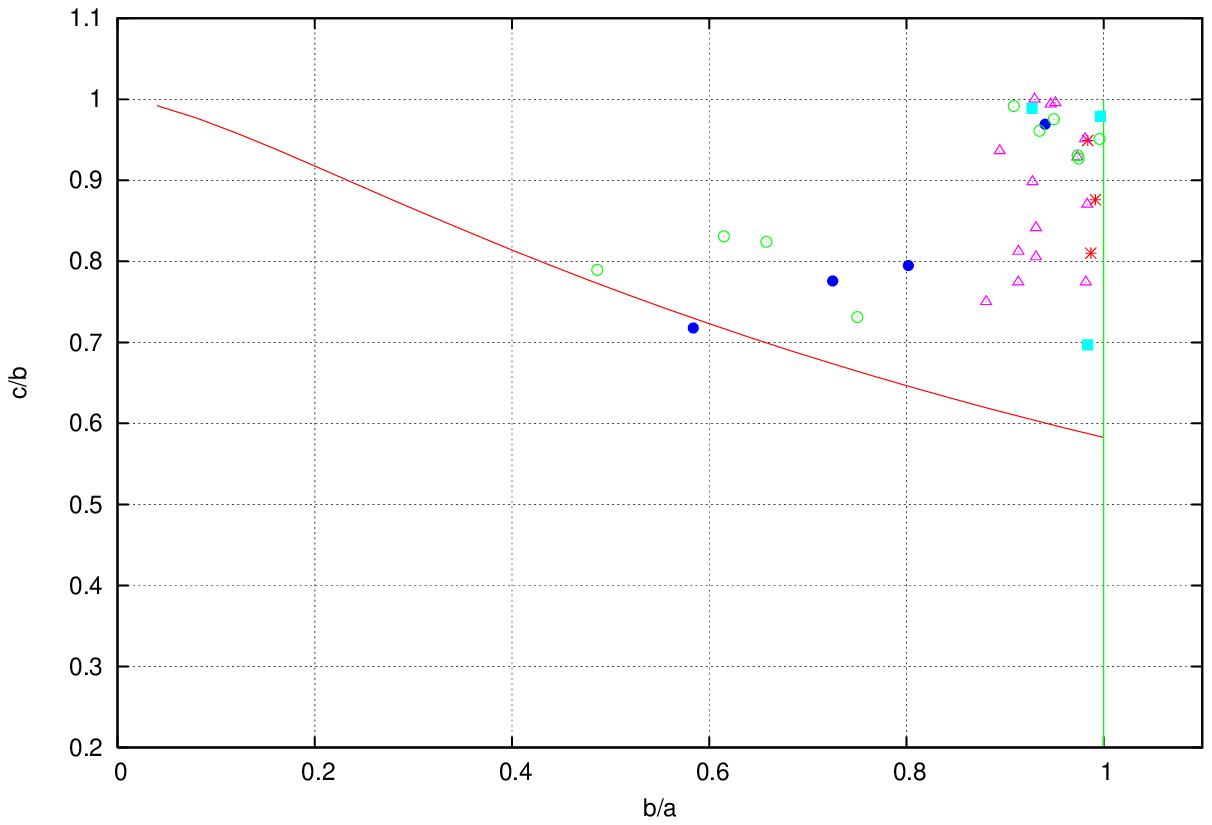}
   }
   \caption{Rapports d'axes obtenus par observation photométrique (à gauche) et des corps 
   formés au cours des simulations numériques (à droite).
\textit{Triangles violets} : résultats de l'effondrement gravitationnel d'un
nuage sphérique ($d_{x}=d_{y}=d_{z}$), \textit{carrés bleus turquoises} : même chose pour un nuage aplati ($d_{x}=d_{y}$), \textit{cercles verts} : même chose pour un nuage aplati ($d_{y}=d_{z}$), \textit{étoiles rouges} : effondrement d'un nuage sphérique mais pour $\rho_{p}\neq 1$g cm$^{-3}$, \textit{disques bleus} : cas de formation de satellites. La séqence de Maclaurin et de Jacobi sont représentées, respectivement, en \textit{vert} et en \textit{rouge}.
   \label{compare}}
\end{figure}

Tout ellipsoïde est défini par ses 3 demi-axes $a>b>c$ ou par ses rapports
d'axes (généralement $b/a$ et $c/a$). Selon certaines études
\cite{Magnus},\cite{Kaasa} les courbes de lumière observées seraient
compatibles avec des formes ellipsoïdales à 3 axes et, plus
généralement, avec des formes convexes en rotation uniforme. 
Récemment, ces résultats ont été renforcé pour certains objets grâce à des
observations menées avec le télescope spatial
Hubble \cite{Hestro1},\cite{Tanga}. Ceci n'exclut pas le fait que les courbes
de lumière puissent contenir, au moins dans quelques cas, beaucoup plus de
renseignements sur des formes davantage complexes \cite{Kaasa}.\\
\\
Il y a une vingtaine d'années, certaines études \cite{Farin2},\cite{Weid}
tentèrent de modéliser les astéroïdes ré-accumulés, ou ``\textit{rubble
  pile}", par des fluides en équilibre hydrostatique. 
Cependant, l'approche ``fluide", qui consiste à calculer la surface
équipotentielle pour un fluide en rotation, comporte certaines faiblesses. Si
l'on considère les rapports d'axes des astéroïdes observés par photométrie, on
s'aperçoit (fig.~\ref{compare}) que peu d'entre-eux possèdent des formes d'équilibre
de fluides \cite{Drum},\cite{Magnus}. 
Cependant, parmi les astéroïdes les plus proches aux formes d'équilibres de fluides, on s'est 
aperçu \cite{Hestro2} que les astéroïdes binaires étaient
particulièrement nombreux. La réciproque, par contre, n'est pas vraie
\cite{Hestro2} : les astéroïdes binaires connus ne possèdent pas tous des 
formes d'équilibres de fluides.
Dans les années 80-90, on s'intéressa davantage aux solutions d'équilibres dans le cas
de fluides compressibles, c'est-à-dire des fluides dont l'indice polytropique 
$n$ est tel que $n \not= 0$ \cite{Hachi},\cite{Lai}.
Plus récemment, il a été montré \cite{Hol} qu'un faible frottement interne suffisait à
rendre compte de quasiment toutes les formes d'astéroïdes observées.\\
Nous reviendrons sur ces modèles (2.3) après avoir montré les éléments qui
conduisent au calcul des formes d'équilibre de fluides (2.2).

\subsection{Généralités et rotation limite}

Tout ellipsoïde triaxial est une surface quadratique dont l'enveloppe 
est donnée, en coordonnées cartésiennes, par :
\begin{equation}
\left(\frac{x}{a}\right)^{2}+\left(\frac{y}{b}\right)^{2}+\left(\frac{z}{c}\right)^{2}=1
\end{equation}
Le triplet $(a,b,c)$ représente, respectivement, les demi-axes de l'ellipsoïde
: grand, moyen et petit. L'ellipsoïde devient 
une sphère si ses trois demi-axes sont égaux. Si deux des trois demi-axes sont
égaux, l'ellipsoïde est appelé sphéroïde (ou ellipsoïde biaxial). Selon si
$c<a$ ou $c>a$, il s'agira respectivement d'un obloïde ou d'un proloïde.\\
On définit ensuite les rapports d'axes $b/a$ et $c/a$.
Sauf mention contraire, l'aplatissement de tout ellipsoïde sera 
défini par $e=\sqrt{1-(c/a)^{2}}$.\\
\\
Quasiment dépourvus de cohésion interne, les agrégats sont 
très sensibles aux forces centrifuges inhérentes à la rotation. Dès lors, on
peut en effet se demander quelle est la vitesse de rotation limite $\Omega_{max}$ à partir de 
laquelle une particule située en périphérie d'un objet sphérique (et plus 
généralement ellipsoïdal) peut s'échapper sous l'effet de la seule force
centrifuge. 
En égalant gravitation et forces centrifuges, on obtient dans le cas d'une sphère la
relation suivante \cite{Rich}:
\begin{equation}
\frac{\Omega_{max}}{\sqrt{2 \pi G \rho_{rp}}}=\sqrt{\frac{2}{3}}
\end{equation}
où $\rho_{rp}$ est la densité du corps considéré. Pour un proloïde, cette relation
prend la forme suivante \cite{Rich} :
\begin{equation}
\frac{\Omega_{max}}{\sqrt{2 \pi G \rho_{rp}}}=\exp^{-3/2}(\bar{e}^{2}-1)^{-1/2}\left[2\bar{e}+\ln\left(\frac{1-\bar{e}}{1+\bar{e}}\right)\right]
\end{equation}
avec ici l'aplatissement $\bar{e}$ est défini par $\bar{e}=\sqrt{1-q^{2}_{2}} \neq e$. Numériquement,
il est parfois utile d'assimiler le proloïde à un corps ponctuel de même masse. On
aboutit alors à la relation simplifiée \cite{Rich} :
\begin{equation}
\frac{\Omega_{max}}{\sqrt{2 \pi G \rho_{rp}}}=\sqrt{2/3}(1-\bar{e}^{2})^{-1/2}
\end{equation}

\subsection{Equilibre hydrostatique}

\begin{figure}
   \resizebox{\hsize}{!}{
     \includegraphics[width=100mm]{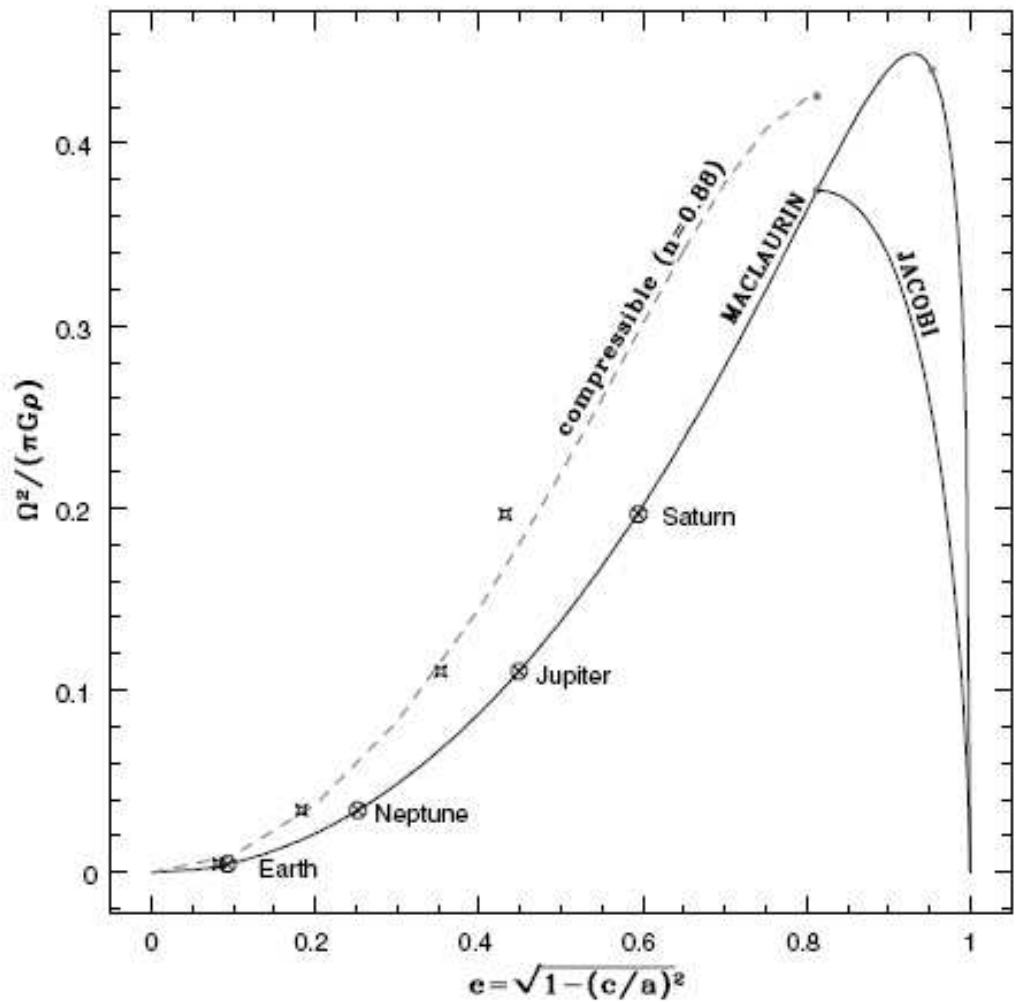}
   }
   \caption{Courbes de Maclaurin et Jacobi \textit{(traits pleins)}. La partie
     de la courbe de Maclaurin située à droite du point de bifurcation
     (Maclaurin - Jacobi) correspond à des états d'équilibres instables. Les états d'équilibres
     d'un fluide compressible ($n$ est l'indice polytropique) sont également
     réprésentés \textit{(tirets)}. L'aplatissement théorique des planètes \textit{(ronds)}, supposées
     homogènes et incompressibles, est supérieur aux valeurs observées \textit{(croix)}.
     D'après \cite{Hestrofig}.
     }
   \label{mclaurinplanets}
\end{figure}

Toute masse fluide, en rotation rigide, soumise uniquement aux forces gravitationnelles
et de pression, suit l'équation de l'équilibre hydrostatique :
\begin{equation}
\rho\frac{Du}{Dt}=-\nabla p+\rho g+\frac{1}{2}\rho\nabla|\Omega\times
x|^{2}+2\rho u\times\Omega
\end{equation}
où $\frac{D}{Dt}=\frac{\partial}{\partial t}+(u.\nabla)$,
$u$ est la vitesse au point $x$, $\rho$ la pression, $g$ l'accélération de
la pesanteur et $\Omega$ la vitesse angulaire.
Dus à la rotation, les deux derniers termes de l'équation sont respectivement la force centrifuge
et la force de Coriolis.
\\
Pour un sphéroïde de Maclaurin ($a=b$) on peut montrer \cite{Chandra} qu'il existe une relation liant
l'aplatissement $e$ du sphéroïde à sa vitesse angulaire $\Omega$ :
\begin{equation}
\frac{\Omega^{2}}{\pi G\rho}=2(3-2e^{2})\frac{\sqrt{1-e^{2}}}{e^{3}}\arcsin(e)-\frac{6(1-e^{2})}{e^{2}}
\end{equation}
De même, il est possible d'exprimer le moment angulaire $L$ en
fonction de la vitesse de rotation $\Omega$ \cite{Chandra} :
\begin{equation}
\bar{L}=\frac{\sqrt{3}}{5}\left(\frac{a}{R}\right)^{2}\bar{\Omega}
\end{equation}
avec $\bar{L}=L/(Gm^{3}R)$ le moment angulaire adimensioné, $R=(a^{2}c)^{1/3}$ le rayon d'une sphère
de même masse $m$ que le sphéroïde et $\bar{\Omega}=\Omega/(\pi G \rho)^{1/2}$
la vitesse de rotation adimensionée.\\
\\
Pour un ellipsoïde de Jacobi ($a \neq b$), on obtient grâce à une algèbre plus compliquée \cite{Chandra} :
\begin{align}
\frac{\Omega^{2}}{\pi G\rho}&=2abc\int_{0}^{+\infty}\frac{u\,du}{(a^{2}+u)(b^{2}+u)\Delta}\\
\Delta^{2}&=(a^{2}+u)(b^{2}+u)(c^{2}+u)
\end{align}
Là encore, on peut établir une relation liant moment angulaire $L$ et
vitesse de rotation $\Omega$ :
\begin{equation}
\bar{L}=\frac{\sqrt{3}}{10}\frac{a^{2}+b^{2}}{R^{2}}\bar{\Omega}.
\end{equation}
\\
Augmenter la rotation contraint le corps fluide à
prendre une forme de plus en plus aplatie. Pour des faibles vitesses de
rotation, deux formes d'équilibre sont possibles : l'une avec un aplatissement
$e<0.8$ et telle que $a=b$ (sphéroïde ; séquence de Maclaurin), l'autre avec
la même vitesse de rotation mais un aplatissement plus grand ($0.8<e<1$) et
telle que $a \neq b$ (ellipsoïde à 3 axes ; séquence de Jacobi). Celle-ci
possède aussi un plus grand moment d'inertie, donc un moment angulaire $L$
plus élevé. Au-dessus de $\bar{\Omega}_{crit}^{2}$, aucune forme stable
n'existe pour un fluide (fig.~\ref{mclaurinplanets}). 

\subsection{Modèle élasto-plastique}

Bien que l'hypothèse d'un comportement fluide semble valable pour les corps
gazeux du Système solaire, rien n'est moins sûr à propos des petits corps
(astéroïdes) pour
lesquels le comportement fluide est loin d'être garanti. 
A priori séduisante, l'idée d'assimiler les astéroïdes à des corps
linéairement élastiques comporte quelques
inconvénients. En effet, les solutions ``élastiques" reposent sur une hypothèse
forte : l'existence d'un état initial sans contraintes dans le passé évolutif du
corps considéré \cite{Hol}. Or, les états actuels des corps du Système Solaire
sont le résultat d'une évolution mouvementée : collisions, fragmentations,
accumulations et réajustements. Ces phénomènes sont à l'origine de
contraintes résiduelles qui ne peuvent être connues avec précision.\\
Adopter une approche élasto-plastique permet de s'affranchir de la
connaissance du passé de l'objet. En effet, en calculant le chargement maximal
accessible pour un corps donné, on peut déterminer des
solutions d'équilibres limites et ce, en particulier, pour des corps composés
de fragments. D'après des études géologiques, le comportement de
certains matériaux granulaires (roches, sols ou graviers) nécessite des théories
élasto-plastiques et répond au critère de Mohr-Coulomb. Déterminé uniquement
par une cohésion $Y$ et un angle de frottement interne $\phi$, ce critère est
généralement utilisé pour calculer la contrainte maximale qu'un corps peut
supporter avant cassure. Etant symétrique, le tenseur des contraintes $\Sigma=[\sigma_{ij}]$
est diagonalisable et admet pour valeurs propres $\sigma_{1} > \sigma_{2} > \sigma_{3}$,
appelées contraintes principales. Dans le cas où la cohésion est négligeable,
le critère de Mohr-Coulomb s'écrit \cite{Hol} :
\begin{equation}
\tan(\phi) \geq \frac{\sigma_{1}/\sigma_{3}-1}{2\sqrt{\sigma_{1}/\sigma_{3}}}
\end{equation}
L'angle de frottement est alors limité uniquement par les contraintes
principales $\sigma_{1}$ et $\sigma_{3}$. Typiquement, les matériaux
granulaires terrestres sont tels que $\phi \sim 30°$. Ainsi, un tas de sable
aura pour pente maximale $\phi \sim 30°$ alors qu'un fluide ($\phi=0°$) ne formera
aucun tas. Il est relativement fréquent 
d'assimiler un matériau à un solide élasto-plastique sans cohésion interne. Couramment employé en
mécanique des sols, ce modèle semble approprié pour décrire la structure interne
des astéroïdes réaccumulés au sein desquels les forces de cohésion sont négligeables \cite{Hol}.

\section{Méthode numérique}

Afin d'aborder la réaccumulation gravitationnelle d'un système de particules
et la formation d'un ou plusieurs agrégats de fragments, nous avons étudié
l'évolution d'un nuage de particules en imposant une dissipation par impacts
de l'énergie cinétique du système. Cela a été rendu possible par l'utilisation
d'un code $N$-corps hiérarchique s'inspirant fortement de l'algorithme de
Barnes-Hut \cite{Barnes}, {\tt pkdgrav} (\textbf{P}arallel \textbf{K}-\textbf{D} tree \textbf{GRAV}ity code).
Les simulations ont été effectuées sur une station de calcul parallèle - SIVAM II - à
l'Observatoire de la Côte d'Azur. 
Bien qu'effectuées sur une station de calcul
parallèle à quatre processeurs, chaque simulation a nécessité, en règle
générale, plusieurs heures de temps CPU.\\

Contrairement aux codes N-corps classiques, {\tt pkdgrav} a été modifié pour
permettre de
traiter les collisions entre les différentes particules du système. Les
particules sont ainsi considérées non pas comme des entités ponctuelles, mais comme des 
sphères de rayons finis. Lorsque deux particules viennent à se rencontrer
elles ne fusionnent pas mais peuvent rebondir les 
unes sur les autres conformément aux lois gouvernant les collisions
inélastiques entre sphères. \\
Prenons l'exemple d'une particule $P$ dont les composantes 
du vecteur vitesse sont $v_{\parallel}$ et $v_{\perp}$. Lorsque la particule 
$P$ entre en collision avec une particule $P\prime$, celle-ci
se retrouve, après collision, avec deux composantes de vitesse 
$v\prime_{\parallel}=\epsilon_{\parallel}v_{\parallel}$ et $v\prime_{\perp}=\epsilon_{\perp}v_{\perp}$ où 
$\epsilon_{\parallel}$ et $\epsilon_{\perp}$ sont respectivement les
coefficients de restitution parallèle et perpendiculaire associés 
aux collisions inélastiques.\\
\\
Dans toutes les simulations, nous avons utilisé une valeur
$\epsilon_{\perp}=0.8$, ce qui permet une dissipation rapide de l'énergie
cinétique lors des collisions mutuelles entre particules. Le nombre de
particules utilisé ($N=500$) s'avère être un compromis entre temps de calcul
et représentation de la forme du corps final obtenu en fin de simulation.\\
\cite{Rich} a en effet montré dans ses 
simulations qu'une valeur $N>200$ 
n'introduisait aucun biais important dans la représentation (la forme) des objets créés.\\
Une fois le corps formé, le code continue à calculer les collisions mutuelles 
entre particules (même si les sphères sont presque en contacts), rendant ainsi 
la simulation très coûteuse numériquement.

\subsection{Conditions initiales}

Les conditions initiales 
consistent en un nuage de masse $M_{total}$, de $N$ particules
de densité $\rho_{p}$. Sa taille est défini par 3 demi-axes de longueurs
$d_{x}$, $d_{y}$ et $d_{z}$.\\
Pour les simulations numériques présentées ici, nous avons pris
$M_{total}=4\times 10^{12}$ kg. Dans le cas idéalisé où 100\% 
de la masse initialement disponible est réaccrétée pour former 
le corps final, ceci correspond à un rayon des corps réaccumulés de
$R_{rp} \sim$ 1 km. Notons toutefois que 
certaines simulations ont été réalisé avec des densités 
$\rho_{p}>1$ g cm$^{-3}$ (resp. $\rho_{p}$ = 2, 3 et 4 g cm$^{-3}$).\\
Un grand nombre de simulations a été réalisé avec 
$d_{x}=d_{y}=d_{z}=10^{4}m \sim 10R_{rp}$, c'est-à-dire avec un nuage 
initial 10 fois plus grand que le corps final obtenu. Enfin, différents profils de vitesses 
pour les particules du nuage ont été utilisés: un profil uniforme et un profil en 
$1/r$ pour lequel la vitesse des particules diminue vers la périphérie du nuage.\\
Les différentes conditions initiales sont donc caractérisées par 
les paramètres suivants : taille et forme du nuage, distribution 
des vitesses et moment angulaire total.
\\
\\
Une fois initialisé, le nuage de particules est tel que :\\
\\
1) les particules qui le composent occupent des positions aléatoires.\\
2) les vecteurs vitesse des particules ont un module aléatoire 
$v=\eta.v_{max}$, où $\eta$ est une variable aléatoire comprise 
entre 0 et 1, de moyenne 0.5, et distribuée soit uniformément 
soit selon une loi en $1/r$. On a, de plus, $0<v_{max}<k.v_{lib}$ 
où $k<1$ est un paramètre ajustable et $v_{lib}$ est la vitesse 
de libération du nuage pour une particule située à sa périphérie.\\
3) Les directions des vitesses sont ajustées aléatoirement par 
un processus de Monte-Carlo afin d'obtenir la valeur de $L$ souhaitée.\\
\\
On peut donc considérer les vitesses initiales au sein du nuage 
comme la somme de deux composantes : une composante aléatoire et 
une autre composante, systématique, qui détermine la rotation 
d'ensemble du nuage. Ceci permet de sélectionner un ``degré de chaoticité" des 
vitesses initiales, et de reproduire ainsi différentes 
situations rencontrées au cours de simulations de 
réaccumulation, tels quelques cas typiques de familles d'astéroïdes 
comme ``Eunomia" (effondrement rapide) ou ``Koronis" (grande 
chaoticité, effondrement lent) \cite{Michel}.
\\
\section{Résultats et analyse}

\begin{figure}
   \resizebox{\hsize}{!}{
     \includegraphics{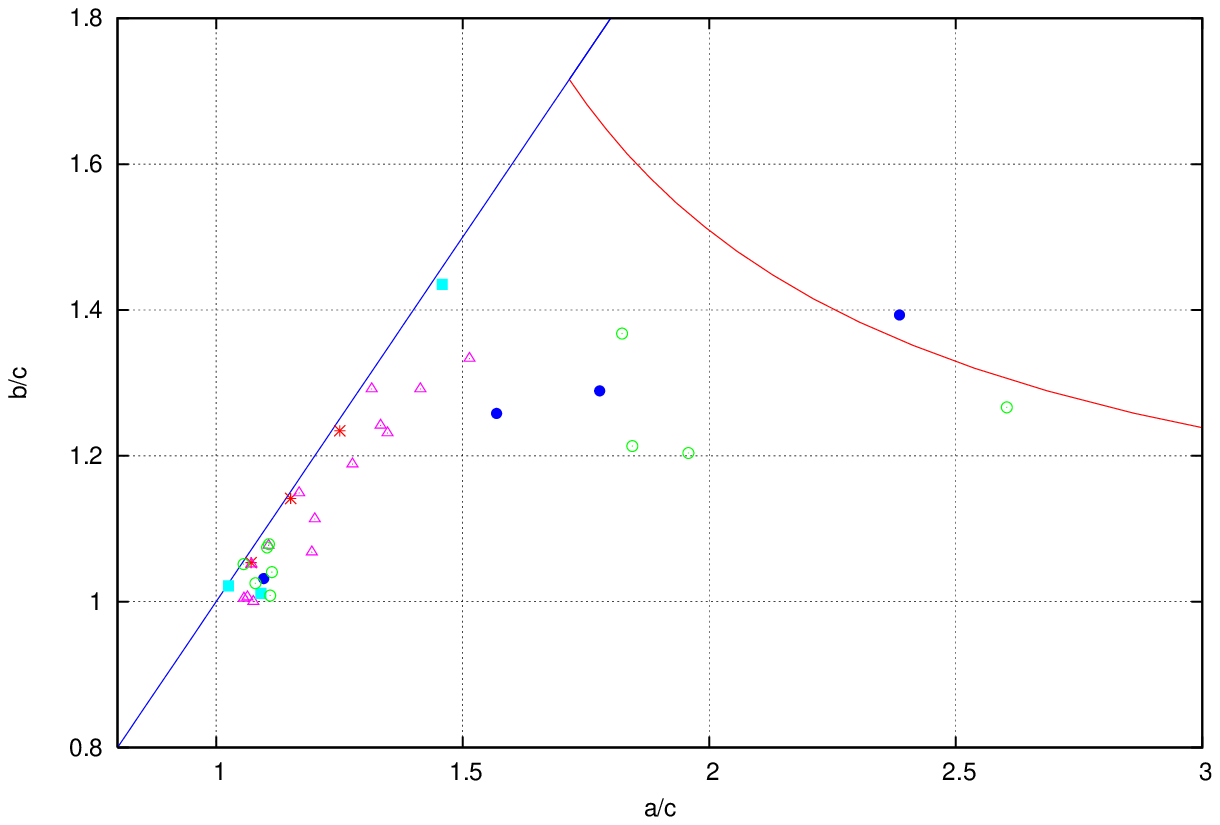}
     \includegraphics{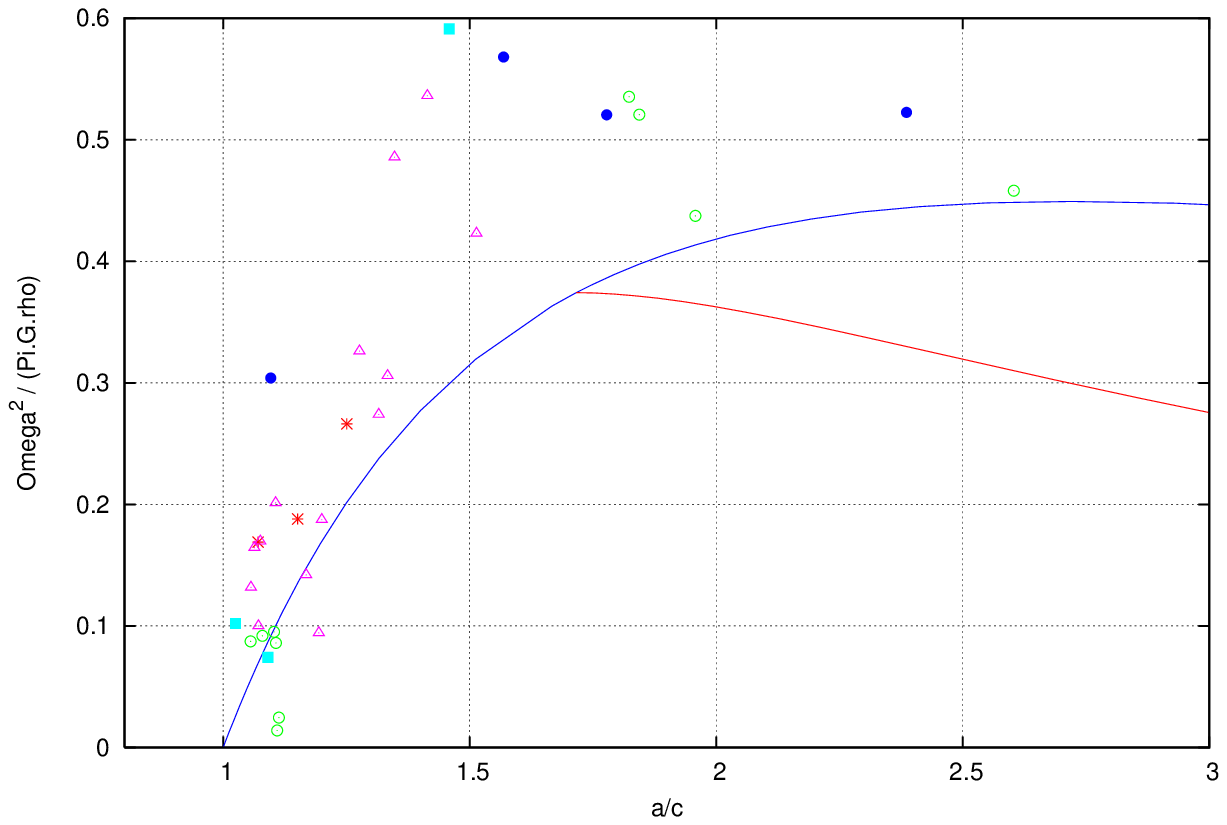}
     \includegraphics{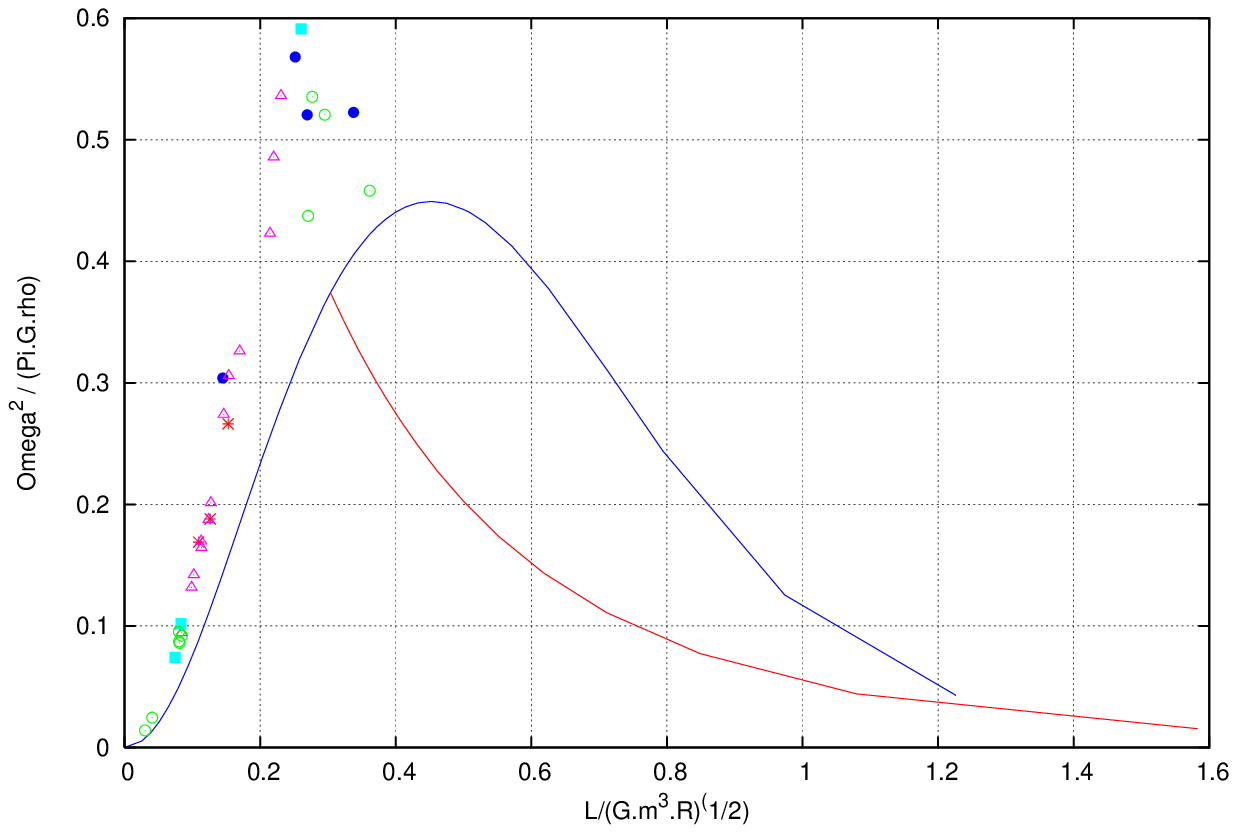}
   }
   \caption{Les differents projections de l'espace des parametres de forme et rotation, defini par 
   $a/c$,$b/c$,$\bar{\Omega}^{2}$,$\bar{L}$.
\textit{Triangles violets} : résultats de l'effondrement gravitationnel d'un nuage sphérique ($d_{x}=d_{y}=d_{z}$), \textit{carrés bleus turquoise} : même chose pour un nuage aplati ($d_{x}=d_{y}$), \textit{cercles verts} : même chose pour un nuage aplati ($d_{y}=d_{z}$), \textit{étoiles rouges} : effondrement d'un nuage sphérique mais avec $\rho_{init}>1$g cm$^{-3}$, \textit{disques bleus} : cas de formation de satellites. La séqence de Maclaurin et de Jacobi sont représentées, respectivement, en \textit{bleu} et en \textit{rouge}.
   \label{resultats}}
\end{figure}

L'agregat principal est identifié grâce à la proximité géométrique des particules.
La vitesse de rotation de l'agrégat est obtenue en calculant le moment 
angulaire total $L$ de l'agrégat, qui lui-même est obtenu en additionnant 
le moment angulaire de chaque particule prise séparément :
$L=\sum_{N}L_{p}$. Les dimensions de l'agrégat sont obtenues en déterminant
les demi-axes $a$, $b$ et $c$. Pour déterminer $a$, on considère la distance entre le 
centre de masse de l'agrégat et le bord 
extérieur de la particule la plus lointaine. Un calcul analogue dans 
les directions perpendiculaires à $a$ fournit successivement une valeur de $b$ et $c$.
Le volume $V_{rp}$ 
de l'agrégat ($V_{rp}=\frac{4}{3}\pi abc$), permet de déterminer 
une valeur de la densité moyenne $\rho_{rp}$ moyenne.
\\
L'ensemble des simulations effectuées est 
présenté dans les différents tableaux (tab. 2 \& 3). 
Les principaux résultats sont représentés sur les figures \ref{compare} et \ref{resultats}. 
Ces différentes simulations ont permis d'examiner, en fonction du moment 
angulaire $L$, les différentes issues possibles du processus de
réaccumulation. 
Nous avons ainsi tenté de vérifier si, pour des valeurs élevées du moment
angulaire $L$, il était possible d'observer la formation 
directe\footnote{Par formation directe, on entend la formation - 
via réaccumulation - de 2 astéroïdes réaccumulés en orbite 
l'un autour de l'autre.} d'astéroïdes binaires via réaccumulation. 
Certaines simulations décrites ci-après se sont achevées par la formation 
de systèmes binaires, confirmant ainsi la possibilité de former de tels 
couples par processus de réaccumulation. 
Les périodes de rotation obtenues sont également en accord avec les périodes
maximales de rotation définies par les relations (2) et (4).

Au cours des simulations, il est apparu que les paramètres 
caractéristiques des corps réaccumulés (masse, rapports d'axes et vitesse angulaire) 
convergent très rapidement, et ne permettent pas d'observer une 
réelle ``relaxation" graduelle. 
Le temps dynamique d'effondrement du nuage initial est donné par 
$t_{eff}=\frac{1}{4}\sqrt{\frac{3\pi}{G\rho_{n}}}$, où 
$\rho_{n}=\frac{M_{tot}}{\frac{4}{3}\pi d_{x}d_{y}d_{z}}$ est 
la densité du nuage initial. L'application numérique donne 
$t_{eff}=0.56$ jours, ce qui est compatible (en ordre de grandeur) 
aux 1.45 jours obtenus expérimentalement.

\subsection{Des sphéroïdes aux ellipsoïdes}

Le phénomène de réaccumulation est par essence très chaotique. Il est 
donc difficile de prévoir a priori la position exacte qu'occupera, en fin 
de simulation, le corps final dans le plan $(\bar{L},\bar{\Omega}^{2})$.\\
Pour des faibles valeurs du moment angulaire ($\bar{L}<$ 0.2), les 
simulations montrent que les corps formés sont typiquement des 
sphéroïdes, c'est-à-dire des corps tels que $a\sim b$ (fig.~\ref{resultats}). De  
formes sphéroïdales, les corps formés se situent au voisinage de la 
séquence de Maclaurin et possèdent des périodes de rotation $P>$ 10 h. 
Ces résultats tendent à montrer que la phase de réaccumulation 
post-collision conduit préférentiellement à la formation de corps 
sphéroïdaux à rotation lente $\sim$ 10 h (relativement lente pour 
des objets d'environ 1 km de rayon). Hormis les cas où l'on 
assiste à la formation de satellites, ces simulations aboutissent 
systématiquement à des corps réaccumulés à plus de 90 \%. Ces corps 
possèdent ainsi des rayons $R \sim $ 1 km et des densités de l'ordre 
de $\rho_{rp}=0.42$ g cm$^{-3}$.\\
Des résultats similaires ont été obtenus en augmentant la densité 
des particules ou, autrement dit, la masse du nuage initial. De 
nouveau, les corps formés en fin de simulation sont proches de 
sphéroïdes. Comme évoqué précédemment, ceci s'explique par le 
fait que la densité $\rho_{p}$ des particules ainsi que la masse 
$M_{total}$ du nuage n'influent pas directement sur les paramètres adimensionés $\bar{L}$ et $\bar{\Omega}^{2}$

Tous les sphéroïdes obtenus se caractérisent par un moment 
angulaire $\bar{L}\leq$ 0.2 et par une vitesse angulaire $\bar{\Omega}^{2}\leq$ 0.3.
Les essais effectués montrent que, pour un moment angulaire $\bar{L}>0.2$, 
les corps formés sont des objets triaxiaux dont la position dans le plan 
$(\bar{L},\bar{\Omega}^{2})$ ne coïncide pas avec les formes d'équilibres 
de Jacobi. On pense \cite{Rich} que les ellipsoïdes ainsi obtenus pourraient,
 sur des échelles de temps plus longues, connaître un ajustement de forme 
ainsi qu'une diminution de leur vitesse de rotation, évoluant ainsi 
progressivement vers des formes d'équilibres de Jacobi.
Au-delà de $\bar{L}\sim 0.2$, l'effondrement gravitationnel entre en 
compétition avec la rotation rapide du nuage. Ainsi, selon la forme 
du nuage et le profil de vitesse utilisé, l'effondrement aura lieu 
uniquement pour la partie centrale du nuage. Un nuage très allongé 
peut alors se fragmenter en sous-ensembles pour donner naissance à 
plusieurs corps pouvant, au gré des interactions, fusionner ou former 
des systèmes binaires ou multiples. Ce scénario semble se vérifier 
dans les simulations : les satellites obtenus sont tous issus de 
l'effondrement de nuages très aplatis, vérifiant 
$d_{x}/d_{y}=d_{x}/d_{z}=16$. Ce comportement, favorale à la formation 
de satellites, est semblable à celui rencontré par Michel et al. \cite{Michel}.

De plus, dans le plan $(\bar{L},\bar{\Omega}^{2})$, deux corps 
pourvus d'un $\bar{\Omega}^{2}$ identique occuperont des 
positions différentes, autrement dit des $\bar{L}$ différents, 
si et seulement si leurs formes respectives sont différentes.\\
En effet, un ellipsoïde tournant autour de son axe d'inertie 
maximale (son petit axe), possède un moment angulaire $L=I\omega$ où 
le moment d'inertie $I$ s'écrit $I=\frac{1}{5}M\left(a^{2}+b^{2}\right)$.\\
Il est important de signaler que les conditions initiales ne 
contiennent aucune ``signature" a priori de la forme finale 
qu'on obtiendra en fin de simulation. En effet, ceci est 
attesté par la variété des conditions imposées qui, pour 
différents moments angulaires $L$, aboutissent à des objets très similaires.\\
Par exemple, afin de ``forcer" le corps à adopter une 
forme différente de la forme sphéroïdale, plusieurs 
simulations ont été réalisés en déformant sensiblement 
le nuage initial : sphérique à l'origine, le nuage a été 
aplati pour devenir ellipsoïdal. Les corps formés au cours 
de ces simulations se situent majoritairement aux environs 
du sommet de la séquence de Maclaurin. Caractérisés par un 
$\bar{L}$ plus élevé que précédemment ($0.2<\bar{L}<0.4$) 
et par $\bar{\Omega}^{2} \geq 0.4$, ces corps possèdent 
des périodes de rotation variant entre $\sim$ 7h et $\sim$ 9h,
cohérent avec les périodes 
observées pour les astéroïdes évoluant dans la Ceinture Principale.\\
\\
D'autre part, ces simulations semblent indiquer qu'un nuage ellipsoïdal, 
doté d'un moment angulaire suffisant, est propice à la formation d'un 
ou plusieurs corps réaccumulés s'organisant autour d'un corps central. 
En effet, on observe dans la zone définie auparavant 
(0.2$<\bar{L}<$0.4 ; $\bar{\Omega}^{2}\geq$0.4) plusieurs de cas de formation de satellites d'astéroïdes.

\subsection{Formation de satellites}

Parmi les simulations effectuées, quatre cas ont abouti à la 
formation de corps secondaires (composés de plus d'une particule) en 
orbite autour d'un corps primaire. Ces corps secondaires ont tous été
 produits à partir de nuages très fortement allongés 
($d_{x}=$ 4$\times$10$^{4}$ m ; $d_{y}=d_{z}=$ 2.5$\times$10$^{3}$ m) 
pour lesquels il a été observé une fragmentation en plusieurs aggrégats 
au cours de l'effondrement gravitationnel. Hormis cinq cas où le 
pourcentage de masse réaccrétée atteint 100\%, toutes les simulations 
montrent des particules en orbite du corps réaccumulé.\\
\\
Parmi ces simulations (tab. 3) :\\
\textbf{HU1 :} formation d'un compagnon (corps secondaire) à partir 
d'une perte de masse du corps primaire. Le rapport de masse secondaire / primaire est tel que $M_{s}/M_{rp}=0.086$). 
A titre de comparaison, on a pour l'astéroïde double Herminone $M_{s}/M_{rp} \sim 0.09$\\
\textbf{H9 :} formation de 2 corps réaccumulés dont l'un 
s'échappe rapidement : absence de satellisation. Le rapport 
de masse secondaire / primaire est faible : $M_{s}/M_{rp}=0.008$.\\
\textbf{H11 :} formation de 3 corps réaccumulés. Alors 
que l'un des 3 est éjecté, les 2 autres se mettent en 
orbite l'un autour de l'autre. Avec un rapport $M_{s}/M_{rp}=0.859$, ce couple
 fait fortement penser à l'astéroïde binaire Antiope dont les deux composantes sont comparables en taille et en masse.\\
\textbf{H12 :} formation de 2 corps réaccumulés en 
orbite l'un autour de l'autre. Ici encore, avec un rapport $M_{s}/M_{rp}=0.792$, ce système binaire obtenu numériquement rappelle l'astéroïde Antiope.

\section{Discussion}

Les formes d'équilibres obtenues dans l'étude de Richardson et al. \cite{Rich} 
occupent une bonne partie du plan $(\bar{L},\bar{\Omega}^{2})$ et se 
situent toutes sans exception à l'intérieur d'un domaine de stabilité 
défini par la relation (3). Toutes ces formes ont également pour 
particularité de se situer en-deça de la limite théorique calculé 
par Holsapple \cite{Hol} pour un angle de frottement $\phi=40°$ 
ce qui renforce encore l'idée que les astéroïdes 
possèdent une structure de type ``agglomérée", en opposition à 
une structure entièrement monolithique.\\
De notre côté, nous observons une variété de formes possibles 
davantage réduite par rapport à \cite{Rich}. Cela signifierait 
donc que, si en effet une grande variété de formes est possible 
(ce qui a déjà été démontré par la stabilité des corps formés 
par \cite{Rich}), seul un sous-ensemble d'entre elles peuvent 
être le résultat de la seule réaccumulation gravitationnelle. 
Les simulations effectuées suggèrent en effet que le processus 
de réaccumulation ne suffit pas à reproduire la grande variété 
de formes réellement observée dans la population d'astéroïdes.
Cependant, nos simulations ont permis de montrer que des 
satellites sont susceptibles d'être créés lors de ces phases 
de réaccumulation gravitationnelle, confirmant ainsi ce qui avait 
déjà été observé dans Michel et al. \cite{Michel} avec une 
utilisation différente du code numérique utilisé ici. Les 
satellites formés au cours de nos simulations résultent d'une 
fragmentation en plusieurs agrégats au cours l'effondrement 
du nuage. Il n'a pas été possible, en revanche, d'observer la 
formation de satellites via la fission d'un corps réaccumulé.

En effet, lorsque le moment angulaire $L$ devient trop élevé, 
la théorie des formes d'équilibre prévoit, comme solution stable, 
un objet binaire en orbite serrée, formé de deux composantes 
identiques. Dès lors, on pourrait s'attendre à observer un gros 
corps qui se divise, ou une accumulation aboutissant directement 
à deux composantes.\\
On pense que le fait de n'avoir pas pu observer de fission est 
une conséquence de la démarche utilisée pour mener les simulations. 
Une fois le corps formé et stabilisé, à la suite de l'effondrement 
gravitationnel du nuage, il n'y a \textit{a priori} aucune 
raison pour que son moment angulaire augmente soudainement 
pour entraîner la fission du corps. Un tel phénomène est plus 
susceptible d'être observé en adoptant une démarche inverse 
à la notre : appliquer des vitesses de rotation croissantes à 
un corps réaccumulé déjà formé. De plus, l'orbite des satellites 
obtenus au cours des simulations n'est pas définitive. Forces de 
marées, champ de gravité du Soleil sont en effet autant de 
facteurs extérieurs pouvant influencer l'évolution de l'orbite 
des satellites sur des échelles de temps beaucoup plus longues 
que la simulation en elle-même.\\
Toutefois, il a été récemment rémarqué \cite{Hestro2} que les 
astéroïdes les plus proches de la séquence d'équilibre fluide 
étaient en réalité des systèmes binaires. Parmi ces corps 
primaires, aucun n'est un ellipsoïde de Jacobi (à 3 axes), 
alors que tous se sont révélés être proches de sphéroïdes 
aplatis (ellipsoïdes à 2 axes ou sphéroïdes de Maclaurin).\\

Un résultat analogue se dégage de nos simulations d'effondrement 
gravitationnel : il semble en effet très difficile d'obtenir des
 corps triaxiaux alors que, par opposition, les simulations forment 
presque exclusivement des sphéroïdes aplatis quelque soit le moment 
angulaire $L$ du nuage initial. Comme évoqué en début de partie, ce 
résultat semble indiquer que les astéroïdes triaxiaux que l'on 
observe ne peuvent être formés directement par réaccumulation 
gravitationnelle. En effet, tout astéroïde subit au cours du temps 
de nombreux impacts mineurs, certes insuffisants pour être détruit, 
mais dont on pense qu'ils seraient suffisamment énergétiques pour 
modifier sensiblement leur forme ainsi que leur vitesse de rotation, 
leur procurant ainsi une position différente dans le plan $(\bar{L},\bar{\Omega}^{2})$.\\
Les sphéroïdes aplatis constitueraient ainsi une catégorie 
de formes dites ``primordiales" qui, au gré d'éventuels impacts, 
pourraient se remodeler et évoluer vers des formes triaxiales, 
tels certains astéroïdes que l'on observe. Il convient toutefois 
de noter que ce mécanisme de modification des formes par impacts 
successifs n'a pas encore été étudié. Il sera donc intéressant 
d'analyser les formes observées par photométrie lors des 
prochaines grandes campagnes d'observations (Gaia, Pan-STARRS),
 car selon des études récentes \cite{Bottke} une partie des 
objets d'un diamètre supérieur à 20 km pourrait être des 
objets dits ``primordiaux", c'est-à-dire n'ayant pas subi d'impacts très 
énergétiques depuis leur formation.\\
Plus généralement, s'il s'avère que les formes biaxiales 
dominent dans la formation par réaccumulation gravitationnelle, 
on peut alors s'attendre à ce que le pourcentage de sphéroïdes 
soit plus important parmi les objets dont la fréquence d'impact 
est plus petite. Une telle vérification 
sera possible lorsque les données seront en nombre suffisant. D'autre part, 
il est tout aussi intéressant de noter que les corps binaires formés par 
simulations se trouvent dans la même région du plan 
$(\bar{L},\bar{\Omega}^{2})$ que les corps binaires observés. 
Une telle observation suggère en effet que les corps binaires observés 
pourraient s'être formés par un processus similaire à la réaccumulation gravitationnelle.

\section{Conclusion}
\label{Conclusion}

Au cours de ce travail, nous avons montré que, même si une
grande variété de formes peut exister pour des agrégats auto-gravitants, 
il est possible que seulement une partie d'entre elles soient accessibles 
par réaccumulation gravitationnelle. La formation directe de corps avec 
un moment d'inertie élevé semble ne pas être favorisée, voire même être totalement inacessible.\\
Les limitations du modèle 
numérique (particules sphériques et de mêmes tailles) ne peuvent 
toutefois être négligées et rendent nécessaires des tests ultérieurs, 
effectués par exemple avec d'autres choix sur la distribution de tailles, 
de vitesses initiales, du nombre de particules, etc. De plus, 
il faut remarquer que tous les éventuels phénomènes agissant 
sur des échelles de temps plus longues (comme les forces de marées) 
ne sont pas pris en compte dans les simulations en raison des 
contraintes liées à leurs durées.\\
Nous avons également 
confirmé que la présence de petits satellites et d'astéroïdes 
binaires (composantes de tailles similaires) était une issue 
possible du processus de réaccumulation, essentiellement pour 
des valeurs élevées de moment angulaire.


\clearpage


\begin{table*}
\caption{Paramètres communs à toutes les simulations. $\delta$ : pas ; $nSteps$ : nombre de pas ; 
$\epsilon_{\perp}$ : restitution perpendiculaire ; $\epsilon_{\parallel}$ : restitution parallèle ; 
$M_{total}$ : masse du nuage initial ; $\rho_{p}$ : densité des particules ; $N_{tot}$ : nombre de 
particules constituant le nuage.
}
\label{table_1}
\centering
\begin{tabular}{cccccccc}
\hline
\hline
$\delta$ & $nSteps$ & $\epsilon_{\perp}$ & $\epsilon_{\parallel}$ & $M_{total}$ & $\rho_{p}$ & $N_{tot}$\\
 (année/2$\pi$) &  &  &  & (kg) & (kg.m$^{-3}$) &  &\\
\hline
\\
2.10$^{-6}$ & 200000 & 0.8 & 1.0 & 4.10$^{12}$ & 1000 & 500\\
\hline
\end{tabular}
\end{table*}

\setlength\tabcolsep{1pt}
\begin{table*}
\caption{Simulations conduisant à la formation d'un corps unique (sans satellites). $L$ est le moment 
angulaire (unité : M$_{\odot}$~UA$^{2}$~(années/2$\pi$)$^{-1}$); $N_{rp}/N_{tot}$ la fraction de 
particules composant le corps réaccumulé; $b/a$ et $c/a$ les rapports d'axes; $e$ l'aplatissement; 
$\rho_{rp}$ la densité et $P$ la période de rotation du corps formé. Les simulation De1, De2, De3 ont été réalisée avec $\rho_{p}$ = 2 g cm$^{-3}$, $\rho_{p}$ = 3 g cm$^{-3}$, $\rho_{p}$ = 4 g cm$^{-3}$}.
\label{table_2}
\centering
\begin{tabular}{ccccccccc}
\hline
\hline
Name & Taille Nuage & $L$ & $N_{rp}/N_{tot}$ & $b/a$ & $c/a$ & $e$ & $\rho_{rp}$ & $P$\\
(profil) & ($d_{x},d_{y},d_{z}$ $\times$ 10$^{4}~$~m) & ($\times$ 10$^{-32}$) &   &   &   &   & (kg.m$^{-3}$) &\\
\hline
H3 (U) & 4,2,2 & 0.733 & 95.0\% & 0.909 & 0.901 & 0.433 & 443.6 & 2.01 j\\
H16 (1/R) & 4,0.25,0.25 & 1.084 & 100\% & 0.935 & 0.899 & 0.439 & 433.6 & 1.54 j\\
H1 (U) & 4,4,2 & 1.908 & 98.0\% & 0.927 & 0.917 & 0.399 & 438.6 & 0.88 j\\
H4 (1/R) & 4,2,2 & 2.060 & 97.4\% & 0.975 & 0.903 & 0.429 & 433.4 & 19.74 h\\
H6 (1/R) & 4,1,1 & 2.084 & 98.8\% & 0.996 & 0.947 & 0.321 & 443.0 & 19.39 h\\
H2 (1/R) & 4,4,2 & 2.112 & 96.7\% & 0.997 & 0.976 & 0.219 & 413.6 & 18.56 h\\
H14 (1/R) & 4,0.25,0.25 & 2.145 & 100\% & 0.974 & 0.906 & 0.423 & 410.2 & 19.30 h\\
D4 (1/R) & 1,1,1 & 2.177 & 99.8\% & 0.981 & 0.933 & 0.359 & 399.2 & 1.03 j\\
H5 (U) & 4,1,1 & 2.178 & 99.0\% & 0.949 & 0.926 & 0.377 & 465.1 & 18.44 h\\
D3 (U) & 1-1-1 & 2.241 & 100\% & 0.894 & 0.838 & 0.546 & 429.9 & 18.93 h\\
H18 (U) & 4,4,4 & 2.517 & 70.5\% & 0.931 & 0.783 & 0.621 & 432.2 & 10.15 h\\
D2 (1/R) & 1-1-1 & 2.624 & 99.8\% & 0.951 & 0.947 & 0.322 & 423.4 & 16.14 h\\
D1 (U) & 1-1-1 & 2.725 & 100\% & 0.983 & 0.856 & 0.517 & 421.3 & 15.59 h\\
D5 (U) & 1,1,1 & 2.986 & 99.6\% & 0.946 & 0.940 & 0.340 & 438.6 & 14.19 h\\
D6 (1/R) & 1,1,1 & 2.997 & 99.8\% & 0.930 & 0.930 & 0.368 & 425.7 & 14.18 h \\
D8 (1/R) & 1,1,1 & 3.262 & 99.6\% & 0.928 & 0.833 & 0.553 & 437.8 & 13.30 h\\
D7 (U) & 1,1,1 & 3.360 & 99.8\% & 0.973 & 0.904 & 0.428 & 441.1 & 15.40 h\\
D10 (1/R) & 1,1,1 & 3.891 & 99.4\% & 0.982 & 0.760 & 0.650 & 401.7 & 11.49 h\\
D9 (U) & 1,1,1 & 4.153 & 100\% & 0.931 & 0.750 & 0.661 & 388.9 & 11.05 h\\
H13 (1/R) & 4,4,4 & 4.782 & 85.5\% & 0.913 & 0.707 & 0.707 & 399.7 & 8.23 h\\
H15 (1/R) & 4,4,2 & 5.560 & 87.6\% & 0.984 & 0.686 & 0.728 & 424.1 & 7.61 h\\
D12 (1/R) & 1,1,1 & 5.599 & 99.0\% & 0.881 & 0.661 & 0.751 & 434.2 & 8.89 h\\
D11 (U) & 1,1,1 & 5.862 & 99.6\% & 0.914 & 0.742 & 0.670 & 444.0 & 8.59 h\\
H17 (1/R) & 4,2,2 & 6.117 & 89.6\% & 0.750 & 0.549 & 0.836 & 432.4 & 7.92 h\\
H7 (1/R) & 4,0.25,0.25 & 7.100 & 99.4\% & 0.615 & 0.511 & 0.860 & 436.4 & 8.72 h\\
H8 (1/R) & 4,0.25,0.25 & 7.668 & 98.8\% & 0.658 & 0.542 & 0.840 & 437.9 & 7.98 h\\
H10 (1/R) & 4,0.25,0.25 & 9.506 & 99.2\% & 0.486 & 0.384 & 0.923 & 423.0 & 8.66 h\\
De1 (1/R) & 1,1,1 & 11.342 & 99.2\% & 0.987 & 0.800 & 0.601 & 860.9 & 7.96 h\\
De2 (1/R) & 1,1,1 & 16.969 & 99.8\% & 0.992 & 0.869 & 0.495 & 1344.3 & 7.58 h\\
De3 (1/R) & 1,1,1 & 23.041 & 99.2\% & 0.984 & 0.934 & 0.358 & 1665.3 & 7.19 h\\
\hline
\end{tabular}
\end{table*}


\begin{table*}
\caption{Simulations conduisant à la formation d'un unique corps accompagné d'un ou plusieurs satellites. $L$ est le moment angulaire (unité : M$_{\odot}$.UA$^{2}$.(années/2$\pi$)$^{-1}$), $N_{rp}/N_{tot}$ la fraction de particules composant le corps réaccumulé, $b/a$ et $c/a$ les rapports d'axes, $e$ l'aplatissement, $\rho_{rp}$ la densité et $P$ la période de rotation du corps formé.}
\label{table_3}
\centering
\begin{tabular}{ccccccccccc}
\hline
\hline
Simulation & Taille Nuage & $L$ & $N_{rp}/N_{tot}$ & $b/a$ & $c/a$ & $e$ & $\rho_{rp}$ & $P$ & $N_{s}$ & $M_{s_{1,2}}/M_{rp}$  \\
(profil de vitesses) & ($x,y,z$ $\times$ 10$^{4}$ m) & ($\times$ 10$^{-32}$ &   &   &   &   & (kg.m$^{-3}$) &  &  & \\
\hline
\\
H12 (1/R) & 4,0.25,0.25 & 1.332 & 51.8\% & 0.941 & 0.912 & 0.410 & 349.8 & 11.69 h & 1 & 0.792 \\
H11 (U) & 4,0.25,0.25 & 2.225 & 51.2\% & 0.802 & 0.637 & 0.770 & 394.1 & 8.06 h & 1 & 0.859 ; 0.067 \\
HU1 (U) & 4,0.25,0.25 & 6.260 & 91.8\% & 0.725 & 0.563 & 0.827 & 407.0 & 8.28 h & 2\? & 0.086 \\
H9 (U) & 4,0.25,0.25 & 8.688 & 96.6\% & 0.584 & 0.419 & 0.908 & 370.0 & 8.67 h & 1 & 0.008 \\
\hline
\end{tabular}
\end{table*}

\clearpage

\clearpage
\begin{figure*}
   \resizebox{\hsize}{!}{
     \includegraphics[angle=0]{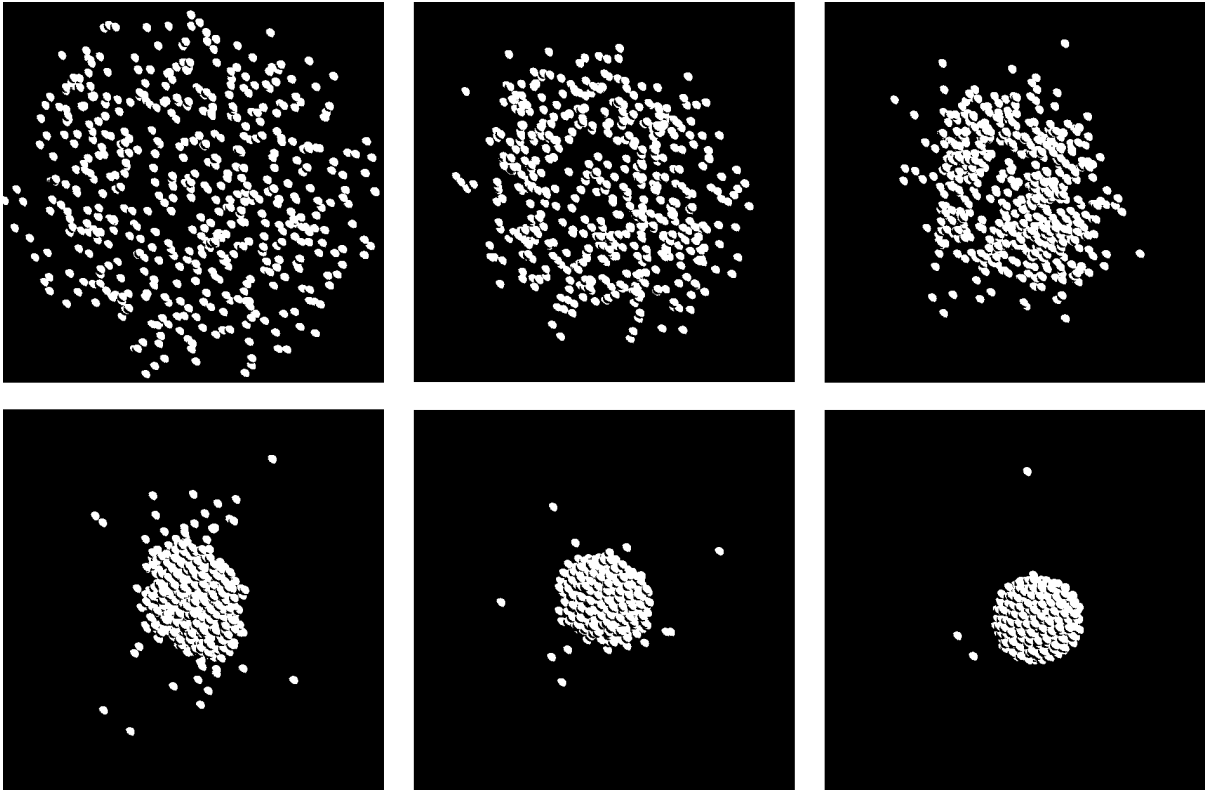}
   }
   \caption{\textit{De gauche à droite, de haut en bas :} effondrement
     gravitationnel d'un nuage sphérique \textit{(simulation D10, cf. tableau
     2)} conduisant à la formation d'un corps ré-accumulé sans
     satellite. Chaque image correspond respectivement à $t=0.01$ jours,
     $t=0.17$ jours, $t=0.23$ jours, $t=0.29$ jours, $t=0.47$ jours et
     $t=23.25$ jours (fin de simulation). L'objet final présente un petit 
nombre de particules-satellites. Sa forme est celle d'un sphéroïde aplati 
dont l'axe de rotation est pratiquement perpendiculaire au plan de la figure.
    \label{simulation}}
\end{figure*}

\clearpage

\clearpage


\begin{thebibliography}{06}

\bibitem[1]{Davis}
D. R. Davis, C.R. Chapman, R. Greenberg, S.J. Weidenschilling, A.W. Harris. Collisional evolution of asteroids : populations, rotations and velocities.
In \textit{Asteroids}, University of Arizona Press, pp. 528-557, 1979.

\bibitem[2]{Farin2}
P. Farinella, P. Paolicchi, V. Zappal\`a. The asteroids as outcomes of catastrophic collisions.
\textit{Icarus} 52, pp. 409-433, 1982.

\bibitem[3]{Durda96}
D. D. Durda. The formation of asteroidal satellites in catastrophic collisions.
\textit{Icarus} 120, pp. 212-219, 1996.

\bibitem[4]{Doress}
A. Doressoundiram, P. Paolicchi, A. Verlicchi, A. Cellino. The formation of binary asteroids as 
outcomes of catastrophic collisions.
\textit{Planetary Space Science} 45, pp. 757-770, 1997.

\bibitem[5]{Michel}
P. Michel, W. Benz, P. Tanga, D.C. Richardson. Formation of asteroid
families by catastrophic disruption : simulations with fragmentation and
gravitationnal reaccumulation. \textit{Icarus} 160, pp. 10-23, 2002.\\
\textbf{Voir aussi :} Collisions and gravitationnal reaccumulation : forming asteroid families and satellites. \textit{Science} 294, pp. 1696-1700, 2001 

\bibitem[6]{Durda03}
D. D. Durda, W.F. Bottke, B.L. Enke, W.J. Merline, E. Asphaug, D.C. Richardson, Z.M. Leinhardt. 
The formation of asteroid satellites in large impacts : results from numerical simulations.
\textit{Icarus} 167, pp. 382-396, 2004.

\bibitem[7]{Chauv}
B. Chauvineau, P. Farinella, F. Mignard. The lifetime of binary asteroids vs. gravitational encounters and collisions
\textit{Icarus} 94, pp. 299-310, 1991.

\bibitem[8]{Schee1}
D. J. Scheeres. Dynamics about uniformly rotating triaxial ellipsoids : applications to asteroids
\textit{Icarus} 110, pp. 225-238, 1994.

\bibitem[9]{Schee2}
D. J. Scheeres. Stability of binary asteroids.
\textit{Icarus} 159, pp. 271-283, 2002.

\bibitem[10]{Pravec}
P. Pravec et 11 collegues. Fast and slow rotation of asteroids.
\textit{Icarus} 148, pp. 12-20, 2000.

\bibitem[11]{Harris}
A. W. Harris. The rotation rates of very small asteroids : evidence for
rubble-pile structure. \textit{Lunar and Planetary Science Conference},
vol. 27, pp. 493-494, 1996.

\bibitem[12]{Vev}
J. Veverka et 16 collegues. NEAR's Flyby of 253 Mathilde : images of a
C-asteroid. \textit{Science} 279, pp. 2109-2114, 1997.

\bibitem[13]{Love}
S.~G. Love, F. H{\"o}rz,  
D.E. Brownlee, Target porosity effects on impact
cratering, and collisional disruption. \textit{Icarus} 105, pp. 216-224, 1993.

\bibitem[14]{Asphaug}
E. Asphaug, S.J. Ostro, R.S. Hudson, D.J. Scheeres, W. Benz. 
Disruption of kilometer-sized asteroids by energetic
collisions. \textit{Nature} 393, pp. 437-440, 1998.

\bibitem[15]{Housen}
K. R. Housen, K.A. Holsapple, M.E. Voss. Compaction as the origin of the unusual craters on the asteroid Mathilde. \textit{Nature} 402,
pp. 155-157, 1999.

\bibitem[16]{Yeo}
D. K. Yeomans et 12 collegues. Estimating the mass of asteroid 253 Mathilde from
tracking data during the NEAR flyby. \textit{Science} 279, pp. 2106-2109, 1997.

\bibitem[17]{Magnus}
P. Magnusson, M.A. Barucci, J.D. Drummond, K. Lumme, S.J. Ostro.
Determination of pole orientations and shapes of asteroids.
In \textit{Asteroids II}, University of Arizona Press, pp. 67-97, 1989.

\bibitem[18]{Kaasa}
M. Kaasalainen, J. Torppa, J. Piironen. Models of twenty asteroids from photometric data.
\textit{Icarus} 159, pp. 369-395, 2002.

\bibitem[19]{Hestro1}
D. Hestroffer, P. Tanga, A. Cellino, F. Guglielmetti, M. Lattanzi,
M. Di Martino, V. Zappal\`a, J. Berthier. Asteroids observations with the Hubble Space 
Telescope FGS : observing strategy, data analysis and modelling process.
\textit{Astronomy \& Astrophysics} 391, pp. 1123-1132, 2002.

\bibitem[20]{Tanga}
P. Tanga, D. Hestroffer, A. Cellino, M. Latttanzi, M. Di Martino, V. Zappal\`a. 
Asteroids observations with the Hubble Space Telescope FGS : duplicity search and size measurements for 6 asteroids. 
\textit{Astronomy \& Astrophysics} 401, pp. 733-741, 2003.

\bibitem[21]{Farin1}
P. Farinella, P. Paolicchi, E.F. Tedesco, V. Zappal\`a. Triaxial equilibrium ellipsoids 
among the asteroids.
\textit{Icarus} 46, pp.114-123, 1981.

\bibitem[22]{Weid}
S. J. Weidenschilling. How fast can an asteroid spin ?
\textit{Icarus} 46, pp.124-126, 1981.

\bibitem[23]{Chandra}
S. Chandrasekhar. \textit{Ellipsoidal figures of equilibrium}.
Yale University Press, 1969.

\bibitem[24]{Hachi}
I. Hachisu, Y. Eriguchi. Fission sequence and equilibrium models of rigidity rotating polytropes.
\textit{Astrophysics and Space Science} 99, pp. 71-74, 1984.

\bibitem[25]{Lai}
D. Lai, F.A. Rasio, S.L. Shapiro. Ellipsoidal figures of equilibrium - compressible models
\textit{Astrophysical Journal Supplement} 88, pp. 205-252, 1993.

\bibitem[26]{Drum}
J. D. Drummond, S.J. Weidenschilling, C.R. Chapman, D.R. Davis. Photometric geodesy of main-belt asteroids : II. Analysis of lightcurves for poles, periods, and shapes.
\textit{Icarus} 76, pp. 19-77, 1988.

\bibitem[27]{Hestro2}
D. Hestroffer, P. Tanga. Figures of equilibrium among binary asteroids.
\textit{AAS, DPS meeting 37}, 15.30, 2005.

\bibitem[28]{Hol}
K. A. Holsapple. Equilibrium configurations of solid cohesionless bodies.
\textit{Icarus} 154, pp. 432-448, 2001.

\bibitem[29]{Rich}
D. C. Richardson, P. Elankumaran, R.E. Sanderson. Numerical experiments with rubble piles : equilibrium shapes and spins.
\textit{Icarus} 173, pp. 349-361, 2005.

\bibitem[30]{Barnes}
J. Barnes, P. Hut. A hierarchical $N \log N$ force-calculation algorithm.
\textit{Nature} 324, pp. 446-449, 1986.

\bibitem[31]{Hestrofig}
D. Hestroffer, P. Tanga. Asteroids from observations to models.
\textit{Lect. Notes Phys.} Series 682, pp. 89-116, 2006. 

\bibitem[32]{Leinfig}
Z. M. Leinhardt, D.C. Richardson. A fast method for finding bound systems in numerical simulations : results from the formation of asteroid binaries.
\textit{Icarus} 176, pp. 432-439, 2005.

\bibitem[33]{Bottke}
W. F. Bottke, D.D. Durda, D. Nesvorn\'y, R. Jedicke, A. Morbidelli, D. Vokrouhlick\'y, H.F. Levison. 
Linking the collisional history of the main asteroid belt to its dynamical excitation and depletion.
\textit{Icarus} 179, pp. 63-94, 2005.


\end{thebibliography}
\end{document}